\documentclass[aps,prc,amsmath,amssymb,aps,superscriptaddress,nofootinbib]{revtex4}
\usepackage{graphicx}
\usepackage{bigints}

\begin{document}
\title{A model for neutrino-nucleus interactions in the GeV region}
\author{M. B. Barbaro}
\affiliation{Department of Physics, University of Turin and INFN, via P. Giuria 1, 10128, Turin, Italy}

\begin{abstract}
We review the recent progress in modelling  neutrino-nucleus scattering, in a framework based on scaling which describes simultaneously the nuclear response to electromagnetic and weak probes.
The study is relevant for the analysis of neutrino oscillation data and the design of the next generation experiments Hyper-Kamiokande and DUNE.
\end{abstract}

\maketitle

The precise knowledge of neutrino-nucleus interactions in the GeV region is crucial for the analysis of long-baseline neutrino oscillation experiments, performed at FermiLab and J-PARC, which aim at improving our knowledge of neutrino properties and, in particular,  at discovering CP violation in the lepton sector \cite{HK,DUNE,FedericoINPC2019}. Being the detectors of these experiments made of complex nuclei, as carbon, argon and oxygen, this task strongly relies on the accurate description of nuclear effects over a wide kinematical domain, going from the quasi-elastic to the deep-inelastic regime.

An important effort has been going on in the last years to construct reliable models, in order to minimise the systematic error related to nuclear physics uncertainties. Conversely, neutrinos can be used to probe nuclear structure and dynamics, and can give complementary information to what can be learnt from electron, photon and hadron scattering reactions.
A comprehensive review of the different approaches to this problem and of the main challenges to be faced in the near future can be found in the NuSTEC White Paper \cite{Alvarez-Ruso:2017oui}. 

The most widely studied reaction up to now, by both experimenters and theorists, is the charged current (CC) inclusive scattering 
\begin{equation}
\nu_l + A \rightarrow l + A' \,,
\end{equation}
where only the outgoing lepton $l$ is detected and a $W$ boson is exchanged. We shall focus on this case, although the understanding of semi-inclusive processes, $\nu_l + A \rightarrow l + (A-1) + N$, in which the outgoing lepton is detected in coincidence with an ejected nucleon (or some other product of the reaction), is extremely important for the correct interpretation of the data.
Semi-inclusive reactions are far more sensitive to the details of the nuclear structure and dynamics than inclusive ones, and allow to discriminate better between different models that might give similar inclusive results. 

From the theoretical point of view, electron and neutrino scattering are strictly related processes: the electromagnetic and weak vector currents are simply related by conservation of the vector current. Validation versus electron scattering data, which are very abundant and  precise, is therefore a necessary benchmark for nuclear models. However, neutrinos probe not only the vector, but also the axial nuclear currents, so that the weak cross section has a richer structure than the electromagnetic one.  This is reflected in the formalism: the double differential cross section is the combination of five independent response functions in the case of neutrino scattering and only two in the case of electron scattering,
\begin{eqnarray}
\left(\frac{d^2 \sigma}{dE' d\Omega}\right)_{\nu_l,l} &=& \sigma_0 \left[V_{CC} R_{CC}^w+V_{CL} R_{CL}^w+V_{LL} R_{LL}^w 
+
V_T R_T^w+V_{T'} R_{T'}^w\right]
\\
\left(\frac{d^2 \sigma}{dE' d\Omega}\right)_{e,e'} &=& \sigma_M \left[v_{L} R_{L}^{em}
+v_T R_T^{em}\right] \,.
\end{eqnarray}
In the above equations the responses $R_K(q,\omega)$ are functions of the momentum $q$ and the energy $\omega$ transferred to the nucleus, $E'$ and $\Omega$ are the energy and scattering angle of the outgoing lepton, $\sigma_0$ and $\sigma_M$ depend on kinematics and couplings, $v_K$ and $V_K$ are functions of the lepton kinematics only (see \cite{Amaro:2004bs} for definitions).

Besides the theoretical differences, the experimental situation in electron and neutrino scattering experiments is not the same: the electron beam energy is usually very well known and this allows to identify different reaction mechanisms by selecting the values of $q$ and  $\omega$. On the contrary, neutrino beams are not monochromatic 
and the measured cross section is the average over a broadly distributed neutrino flux $\phi(E_\nu)$, 
\begin{equation}
\left < \frac{d^2 \sigma}{dE' d\Omega}\right >_{\nu_l,l}  = \frac{\bigintsss{ \left(\frac{d^2 \sigma}{dE' d\Omega}\right)_{\nu_l,l} \phi(E_\nu) dE_\nu}}{\int \phi(E_\nu) dE_\nu} \,.
\end{equation}
 As a consequence, given a certain kinematics of the outgoing lepton,  different mechanisms overlap and cannot be disentangled experimentally. These include quasi-elastic scattering, excitation of two-particle-two-hole (2p2h) states, pion production, resonance excitation, deep inelastic scattering, and should in principle be described  within a consistent framework. This is clearly a very challenging task, and all approaches necessarily involve some approximations. Whatever these approximations are, a reliable model should fullfill some basic requirements. Specifically, it should be:
\begin{enumerate}
\item relativistic, or at least contain relativistic ingredients, being the energies involved typically of the order of, or higher than, the nucleon mass;
\item able to describe electron scattering data;
\item easily implementable in Monte Carlo generators, which play a major role in the experimental analyses \cite{Mosel:2019vhx}.
\end{enumerate}
Guided by the above criteria, a model, indicated as SuSAv2 (SuperScaling Approach version 2), has ben developed in Refs. \cite{Gonzalez-Jimenez:2014eqa,Megias:2016fjk}.  The model is based  on the analogies between electron- and neutrino-scattering and uses the property of scaling and superscaling of inclusive $(e,e')$ data \cite{Donnelly:1999sw} to connect the two processes \cite{Amaro:2004bs}.
The scaling functions, which embody the nuclear dynamics in the different channels (longitudinal, transverse, isoscalar, isovector), are evaluated in the relativistic mean field (RMF) framework. 
Hence the model satisfies the first two above requirements. Moreover, it has recently been implemented in the generator GENIE~\cite{Dolan:2019bxf}, while its implementation in NEUT is in progress.
 
 Let us  outline the main ingredients of the SuSAv2 model and refer  to \cite{Gonzalez-Jimenez:2014eqa,Megias:2016fjk}  for further details:
 \begin{itemize}
 \item[-]
In  the quasi-elastic region  the Impulse Approximation (IA) is used:  the probe is assumed to interact with a single nucleon bound in the nucleus.
\item[-]
 The underlying Lagrangian is  Walecka-type one and the fundamental degrees of freedom are protons and neutrons interacting through the exchange of scalar and vector mesons.
\item [-]
 The nucleon bound states are solutions of the Dirac equation in a finite nucleus in presence of strong, energy-independent, scalar (S) attractive and vector (V) repulsive potentials,  fitted to ground state nuclear properties.
 \item[-]
 The ejected nucleon wave function depends on the treatment of final state interactions (FSI) with the residual nucleus. In the model, these are described by the same S and V potentials used for the bound state. This approach has the merit of  preserving orthogonality between the initial and final wave functions, a property violated by other approaches based on the Plane Wave Impulse Approximation (PWIA) or on optical potentials.
 However,  it has the drawback of giving a too strong mean field at high momentum and energy transfers, when the ejected nucleon should no more feel the residual interaction and is better described by a plane wave. This is a direct consequence of the energy-independence of the potentials.
 In order to correct for this behaviour,  a "blending function", which mixes the PWIA and the RMF approaches depending on the momentum transfer $q$, has been constructed. This introduces two parameters, that are fitted once and for all to electron scattering data on $^{12}$C.
 Recently this empirical recipe has been compared with a microscopic calculation \cite{Gonzalez-Jimenez:2019ejf} in which the final state is described by a relativistic distorted wave stemming from the real part of an optical potential. It has been shown that the two approaches yield very similar results, providing a more solid justification of this prescription.
 \item[-]
  The model includes two-body currents mediated by pions (MEC) \cite{Amaro:2010sd,Amaro:2011aa}. These contributions go beyond the IA and are responsible for the excitation of 2p2h states, typically appearing in the energy spectrum between the QE and the $\Delta$ resonance peaks - the so-called "dip" region. The details of the MEC calculation can be found in  \cite{Simo:2016ikv}.
       \end{itemize}

In order to illustrate the performances of the model, we now show  a few representative results. Interested readers can find many more results  in Refs. \cite{Megias:2016fjk,Megias:2018ujz,Barbaro:2019vsr,Ivanov:2018nlm,Barbaro:2018kxa,Megias:2017cuh}.

\begin{figure}[h]
\includegraphics[width=12pc,angle=-90]{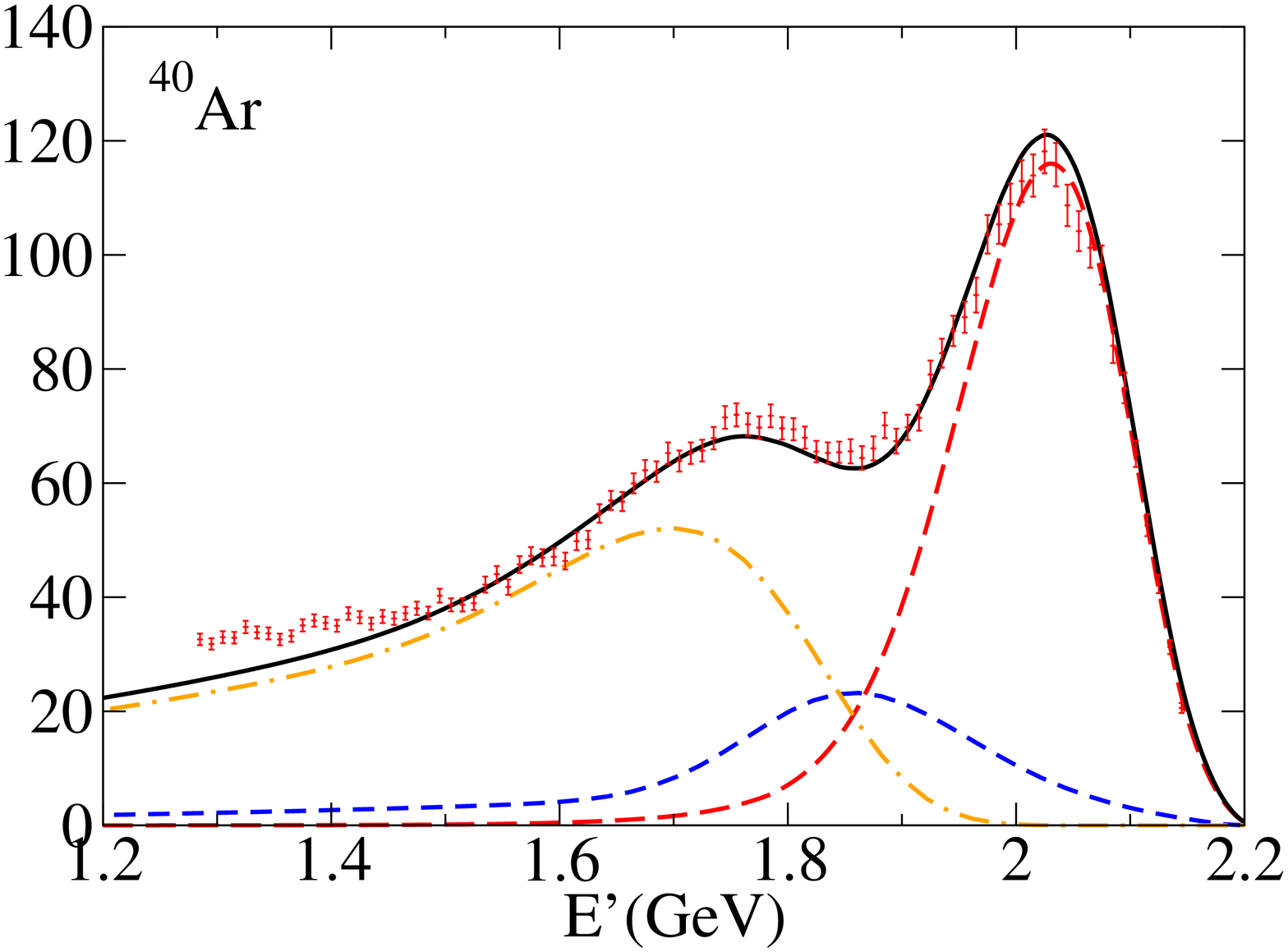}
\hspace{2pc}
\includegraphics[width=12pc,angle=-90]{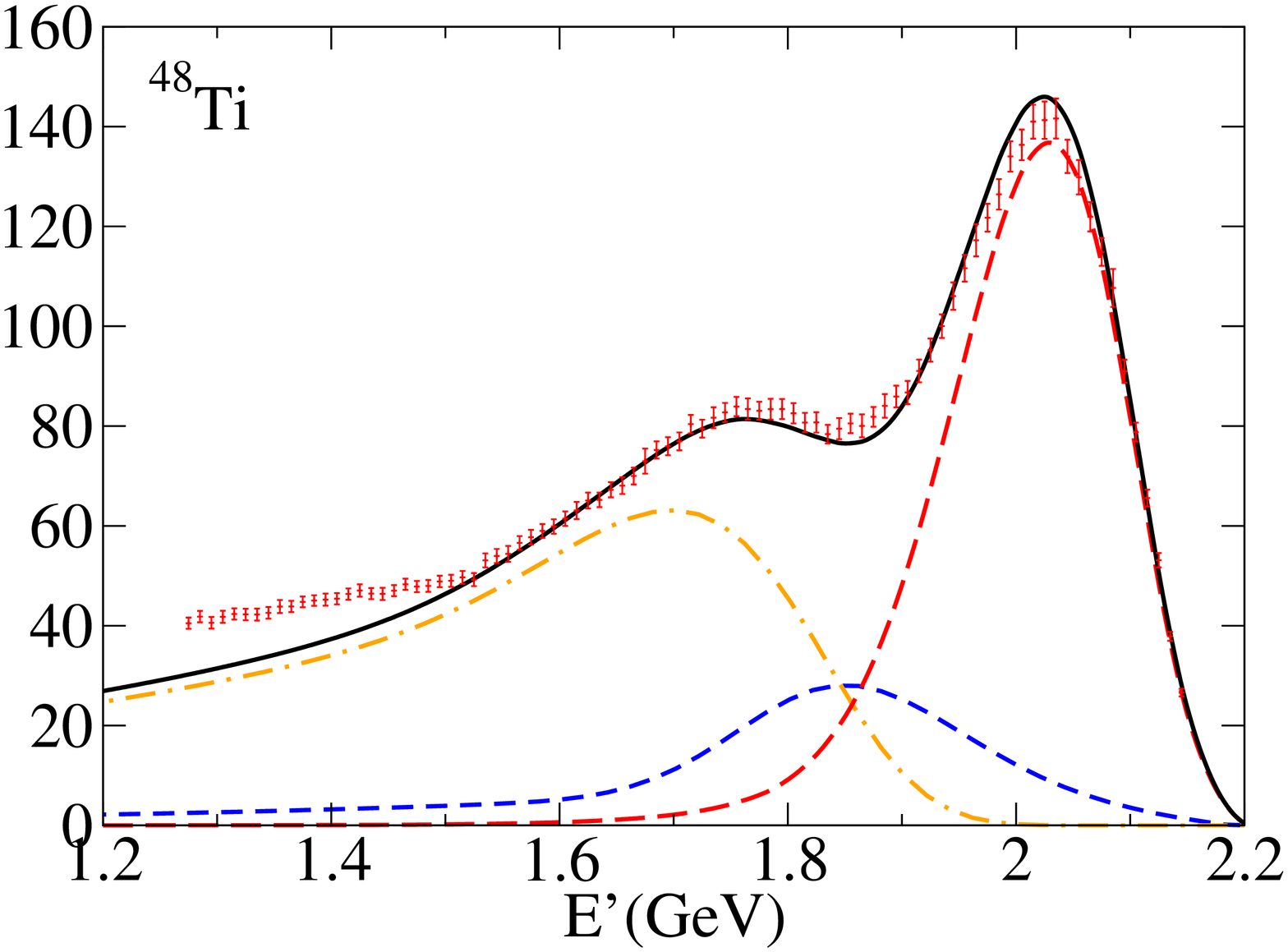}
\caption{\label{fig:Jlab} The $(e,e')$ double differential cross section $d^2\sigma/d\Omega/dE'$ (in $\mu b/sr/GeV$) of argon  and titanium from  \cite{Dai:2018xhi,Dai:2018gch}, compared with the SuSAv2 prediction (black solid curve). The separate QE (red dotted), 2p2h (blue dashed) and inelastic (orange dot-dashed) contributions are also shown. The beam energy is $E$=2.222 GeV and the scattering angle $\theta$=15.541 deg. Figure from \cite{Barbaro:2019vsr}.}
\end{figure}

In Fig.~\ref{fig:Jlab} we present the predictions of the SuSAv2 model compared with the Jefferson Lab electron scattering data on $^{40}$Ar  and $^{48}$Ti  \cite{Dai:2018gch,Dai:2018xhi}. This is an important  test of the model since argon (N=22, Z=18) will be the nuclear target in the DUNE experiment, while the proton distribution in titanium (N=26, Z=22) is expected to be similar to the neutron distribution in argon, which is not directly accessible.
 The agreement between theory and data is excellent over most of the energy spectrum covering the QE, dip and a significant region in the inelastic domain. Work is in progress to understand the discrepancy in the most extreme inelastic regime. Note that no free parameters are used to produce these results, the only parameters of the model being fitted to $(e,e')$ data on $^{12}$C.

\begin{figure}[ht]
\begin{minipage}{18pc}
\includegraphics[width=12pc,angle=-90]{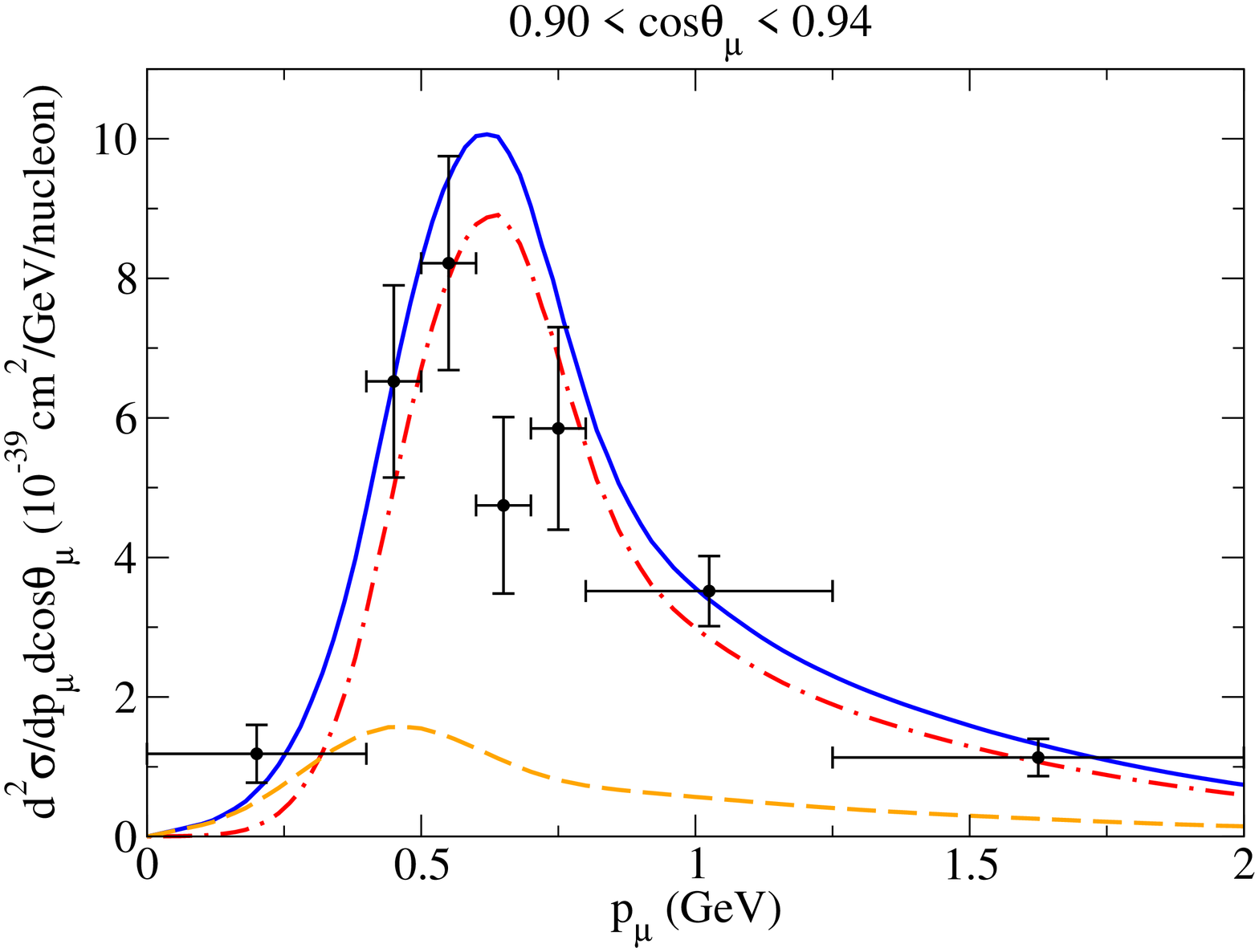}
\caption{\label{fig:T2K}
T2K flux-folded double differential
cross section per target nucleon for the $\nu_\mu$ CCQE process on
$^{12}$C displayed versus the muon momentum  obtained within the SuSAv2 approach. QE
and 2p-2h MEC results are also shown separately. Data are
from \cite{Abe:2016tmq}. Figure from \cite{Megias:2016fjk}.}
\end{minipage}\hspace{3pc}%
\begin{minipage}{18pc}
\includegraphics[width=17pc]{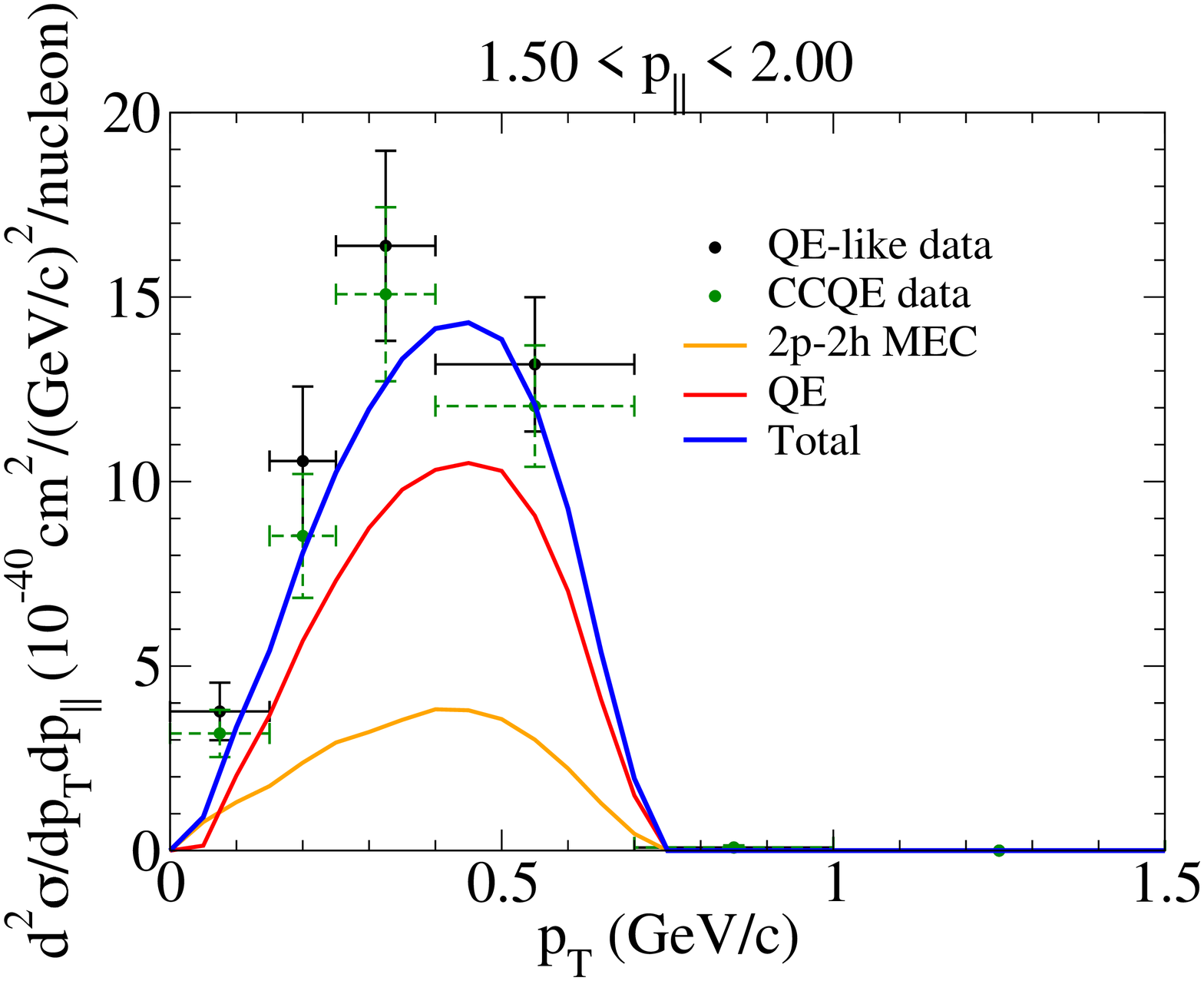}
\caption{\label{fig:Minerva}The MINERvA CCQE double differential cross sections for $\bar\nu_\mu$  scattering on hydrocarbon versus the muon transverse momentum, for a bin of the muon longitudinal momentum (in GeV/c). The  prediction of the SuSAv2 model (blue) is shown as well as the separate QE (red) and 2p2h (orange) contributions.  Data from \cite{Patrick:2018gvi}. Figure from \cite{Megias:2018ujz}.\label{label}}
\end{minipage} 
\end{figure}
 
In Fig.~\ref{fig:T2K} we compare the SuSAv2 prediction with the data of the T2K experiment \cite{Abe:2016tmq}, corresponding to an average neutrino energy $<E>\sim$ 600 MeV. We present the flux-averaged double differential cross section as a function of the muon momentum, for a specific bin in the scattering angle, showing the separate contributions of the pure QE, the 2p-2h, and their sum.
As observed, the model predictions are in general consistent with the error bars,  and the addition of the 2p-2h MEC does not seem to improve in a clear way the comparison with data. The situation is different for the MINERvA experiment, as shown in Fig.~\ref{fig:Minerva} for a set of antineutrino data \cite{Patrick:2018gvi}. In this case, due to the higher antineutrino energy $<E>\sim$ 3.5 GeV and to the larger spreading of the flux, the 2p2h contribution is essential to reproduce the data \cite{Megias:2018ujz}.
A similar level of agreement has been found in the comparison with all other available experimental data.

\vspace{1.cm}

Summarizing, we have developed a nuclear model which  provides very good agreement with inclusive electron scattering data and is capable of reproducing, within the present error bars, the inclusive neutrino and antineutrino data collected  by the MiniBooNE, T2K and MINERvA collaborations. It should be stressed, however, that the agreement with inclusive data is not sufficient to assure the level of control of nuclear effects required by the ambitious program of future neutrino experiments, HyperK and DUNE. Further studies of semi-inclusive reactions, which are more sensitive to the model details, have to be pursued and will likely be one of the top challenges for nuclear theorists working in this field in the next years.

\section*{Acknowledgements}

This contribution is a short and necessarily incomplete summary of a project performed in collaboration with J.E. Amaro, A.N. Antonov, J.A. Caballero, A. De Pace, T.W. Donnelly, R. Gonzalez-Jimenez, M. Ivanov, G.D. Megias, I. Ruiz Simo and J.M. Udias. I would like to thank them all for many stimulating discussions.

\end{document}